\documentclass[aps,twocolumn]{revtex4}%
\usepackage{amsfonts}
\usepackage{amsmath}
\usepackage{amssymb}
\usepackage{graphicx}%
\begin{document}
\preprint{ }
\title[ ]{Extracting Kondo temperature of strongly-correlated systems from the inverse local magnetic susceptibility}
\author{A. A. Katanin}
\affiliation{Center for Photonics and 2D Materials, Moscow Institute of Physics and Technology, Institutsky lane 9, Dolgoprudny, 141700, Moscow region, Russia\\
M. N. Mikheev Institute of Metal Physics, Kovalevskaya str. 18, 620990, Ekaterinburg, Russia}

Arising from X. Deng et al. Nature Communications https://doi.org/10.1038/s41467-019-10257-2 (2019)
\vspace{0.6cm}

\maketitle

The temperature scales of screening of local magnetic and orbital moments are important characteristics of strongly correlated substances.
In a recent paper 
X. Deng et al.  \cite{paper} using dynamic mean-field theory (DMFT) have identified temperature scales of the onset of screening in orbital and spin channels in some correlated metals from the deviation of temperature dependence of local susceptibility from the Curie law. We argue that the scales obtained this way 
are in fact much larger, than the corresponding Kondo temperatures, and, therefore, do not characterize the screening process. 
By reanalyzing the results of this paper we find the characteristic (Kondo) temperatures for screening in the spin channel $T_K\approx 100$~K for V$_2$O$_3$ and $T_K\approx 350$~K for Sr$_2$RuO$_4$, which are almost an order of magnitude smaller than those for the onset of the screening estimated in the paper ($1000$~K and $2300$~K, respectively); for V$_2$O$_3$ the obtained temperature scale $T_K$ is therefore comparable to the temperature of  completion of the screening, $T^{\rm comp}\sim 25$~K, which shows that the screening in this material can be described in terms of a single temperature scale.


X. Deng et al. \cite{paper} have performed a detailed analysis of the temperature dependence of orbital and magnetic local susceptibilities of two strongly correlated materials, Sr$_2$RuO$_4$ and V$_2$O$_3$ within DMFT \cite{DMFT,DMFT2}. At high temperatures the susceptibilities obey the Curie law, $\chi(T) \sim 1/T$. 
The temperatures $T^{\rm ons}$ of the onset of screening of spin- and orbital local moments are obtained from the deviation of 
$T\chi(T)$ from a constant value. Corresponding temperature scales $T^{\rm ons}$ are found to be much larger than the scales, corresponding to the completed screening (onset of the Fermi-liquid behavior) $T^{\rm cmp}\sim 25$~K.

In the following we argue however that the temperatures $T^{\rm ons}$, obtained by the authors, do not correspond to the temperature scales of the spin screening. Indeed, instead of considering the quantity $T\chi(T)$, we plot inverse spin susceptibility $\chi^{-1}(T)$ for both considered compounds, Sr$_2$RuO$_4$ and V$_2$O$_3$ on the basis of the data of the paper (see Fig. 1). For Sr$_2$RuO$_4$ (see Fig. 1a) we do not find any peculiarity at the onset temperature $T^{\rm ons}=2300$~K suggested by the authors. Instead, at all considered temperatures the local susceptibility follows the Curie-Weiss law 
\vspace{-0.2cm}
\begin{equation}
\chi(T)=\frac{C}{T+\theta}
\end{equation}
with a positive temperature $\theta \approx 500K$ (in agreement with earlier result of Ref. \cite{Georges} and experimental data \cite{Sr2RuO4_NMR}). Following Wilson's result for the local spin $S=1/2$ Kondo problem \cite{Wilson,Melnikov,Tsvelik}, the temperature $\theta \simeq \sqrt{2}T_K$ yields the temperature scale of screening of the local moment (Kondo temperature) $T_K$. Since the  dependence $\chi(T)/\chi(0)$ is almost universal for different $S$ \cite{Desgranges}, the abovementioned relation between $\theta$ and $T_K$ is also expected to hold approximately for arbitrary local spin value. Therefore, for Sr$_2$RuO$_4$ we find the temperature scale of spin screening $T_K\approx 350$~K, which is much smaller than $T^{\rm ons}$, obtained by the authors. We also note that very similar linear dependence of the inverse susceptibility is observed in the other Hund metals: $\alpha$-iron ($T_K=50$~K for density-density interaction and $T_K\approx 320$~K for Kanamori interaction)  \cite{a-iron,iron-nickel}, $\gamma$-iron ($T_K\sim 700$~K)  \cite{g-iron}, nickel ($T_K\sim 850$~K)  \cite{iron-nickel}, etc.

\begin{figure*}[t]
\center \includegraphics[width=0.9 \linewidth]{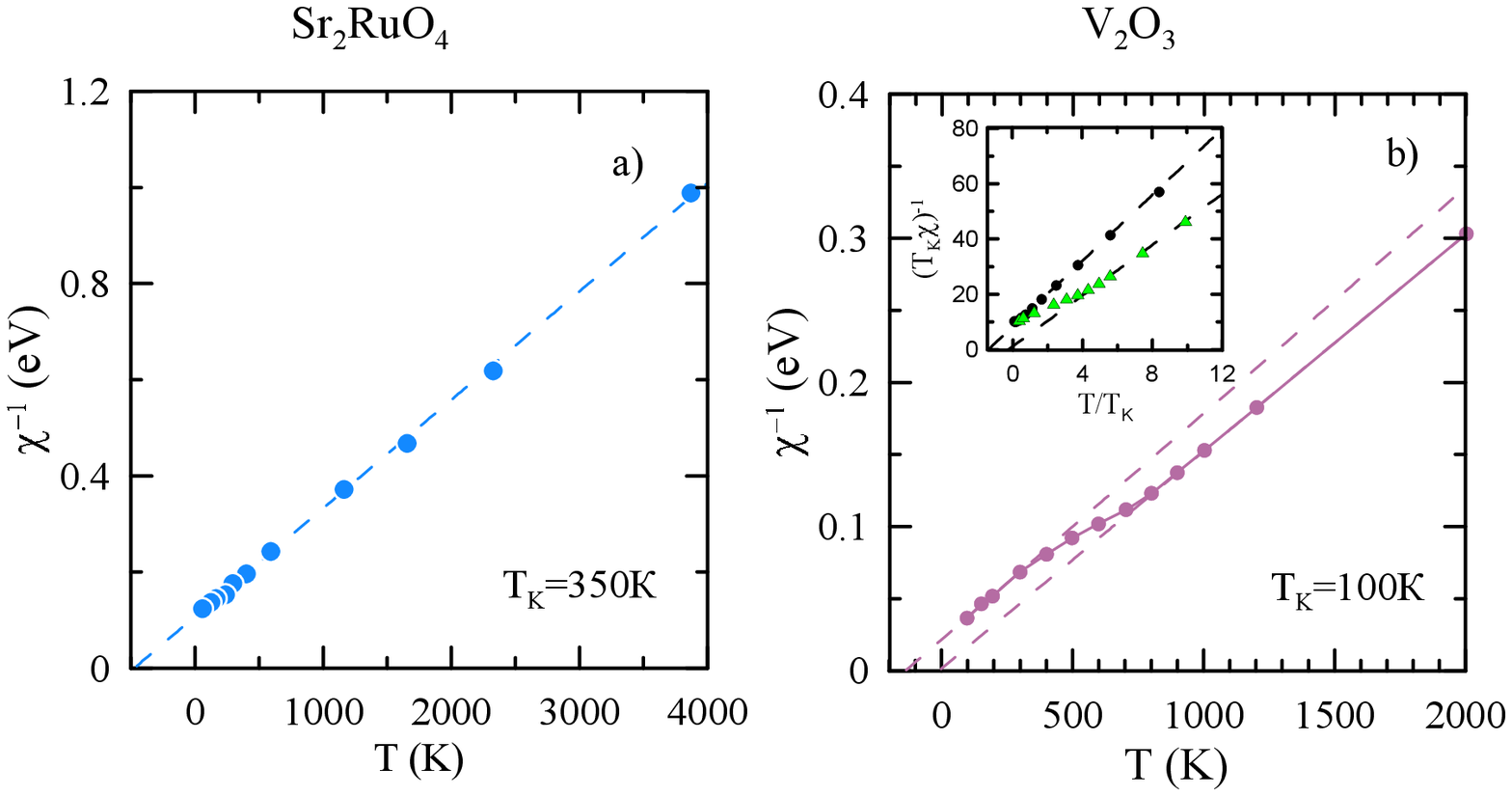} \endcenter
\caption{Temperature dependence of inverse local susceptibility $\chi^{-1}(T)$, calculated from the data of Ref. \cite{paper} for Sr$_2$RuO$_4$ (a) and V$_2$O$_3$ (b). The inset in (b) shows inverse local spin susceptibility of the single-band half filled Hubbard model on the square lattice with nearest neighbour hopping $t$ and on-site Coulomb repulsion $U=9t$ (triangles) in comparison to the inverse spin susceptibility of the Kondo model \cite{Wilson} (circles); the Kondo temperature $T_K=0.032t$ of the Hubbard model is extracted from the fit of the low-temperature part of the susceptibility to the Kondo model.  Dashed lines show linear fits to the data.}%
\label{FigGamma}%
\end{figure*}

For V$_2$O$_3$ the situation is more complex, since the inverse susceptibility shows at $T\sim 600$~K a crossover (see Fig. 1b) from the Curie behavior ($\theta\approx 0$) to Curie-Weiss behavior with $\theta \approx 150$~K. This crossover, however, is likely not related to the screening, but reflects passing from a crossover regime to metallic one in the vicinity of Mott metal-insulator transition for this compound \cite{V2O3}. To confirm this viewpoint, we present in the inset of Fig. 1b the temperature dependence of the inverse local spin susceptibility, obtained for the single-band half filled Hubbard model on the square lattice with nearest neighbor hopping $t$ (on-site Coulomb repulsion $U=9t$ is in the vicinity of Mott transition), showing that this dependence is qualitatively similar to the one, obtained for V$_2$O$_3$.
Therefore, the screening scale of local magnetic moments in V$_2$O$_3$ is again given by the Kondo temperature $T_K=\theta/\sqrt{2}\approx 100$~K, extracted from the low temperature part of the local susceptibility in paramagnetic phase. The latter value is also much smaller than obtained by the authors and has the same order of magnitude as the temperature, at which the screening is completed, $T^{\rm cmp}\sim 25$~K. This makes it reasonable to describe spin screening in V$_2$O$_3$
in terms of a single energy (or temperature) scale, as it should be for a screening process of a single impurity site, considered in DMFT.  We note that rather large Weiss temperature of local susceptibility of V$_2$O$_3$ ($\sim 600K$), observed experimentally in nuclear magnetic resonance studies \cite{V2O3_NMR} in the temperature range $T>150K$ may be related to the impact of strong antiferromagnetic correlations on local susceptibility, which is absent in paramagnetic DMFT solution.

The observation that for V$_2$O$_3$  $T_K\sim T^{\rm cmp}$ is in contrast to the above described situation in Sr$_2$RuO$_4$, where $T_K\gg T^{\rm cmp}\sim 25$~K. We note that such an inequality is also fulfilled for nickel  \cite{iron-nickel}, and in that case this was attributed to underscreened Kondo effect, since the Fermi level of nickel is close to the upper edge of the band. The origin of strong difference of Kondo temperature and the temperature, corresponding to completion of the screening in Sr$_2$RuO$_4$, requires further studies and clarification. 
\\
{\bf Acknowledgements. } The work is partly supported by the theme ``Quant"
AAAA-A18-118020190095-4 of Minobrnauki, Russian Federation.\\
{\bf Data availability.} The DMFT data for susceptibility of Sr$_2$RuO$_4$ and V$_2$O$_3$, analyzed here, are taken from Ref. \cite{paper}. The data on the single-band model are available within the present  paper. \\
{\bf Author contributions} A. K. is the sole author of this work and is responsible for the conception and design of the
work. \\
{\bf Competing interests:} The author declares no competing interests.

\end{document}